\documentclass[aps,amsmath,amssymb,preprint,prd]{revtex4-1}
\usepackage{graphics}      
\usepackage{graphicx}      
\usepackage{longtable}     
\usepackage{url}           
\usepackage{bm}            
\newcommand{\beq}{\begin{equation}}
\newcommand{\eeq}{\end{equation}}
\newcommand{\bea}{\begin{eqnarray}}
\newcommand{\eea}{\end{eqnarray}}
\newcommand{\meio}{{}^1\!/{}_{\!2}}

\begin{document}

\title{Relativistic Landau Levels in the Rotating Cosmic String Spacetime}

\author{M. S. Cunha{${}^{\dagger}$}, C. R. Muniz{${}^{\ddagger}$}, H. R. Christiansen{${}^{\diamondsuit}$}, V. B. Bezerra{${}^{\! *}$}}

\affiliation{${}^\dagger$Grupo de F\'isica Te\'orica (GFT), Universidade Estadual do Cear\'a, 60714-903, Fortaleza-CE, Brazil. \\ ${}^{\ddagger}$\!Universidade Estadual do Cear\'a, Faculdade de Educa\c c\~ao, Ci\^encias e Letras de Iguatu, Rua Deocleciano Lima Verde, Iguatu-CE, Brazil. \\ ${}^{\diamondsuit}$\!Instituto Federal de Ci\^encia, Educa\c{c}\~ao e Tecnologia, IFCE Departamento de F\'isica, 62040-730 Sobral, Brazil.\\ ${}^{\! *}$\!\!\! Departamento de F\'isica, Universidade Federal da Para\'iba-UFPB, Caixa Postal 5008, Jo\~ao Pessoa-PB, 58051-970, Brazil.
}

\begin{abstract}
In the spacetime induced by a \emph{rotating} cosmic string we compute the energy levels of a massive spinless particle coupled covariantly to a homogeneous magnetic field parallel to the string. Afterwards, we consider the addition of a scalar potential with a Coulomb-type and a linear confining term and completely solve the Klein-Gordon equations for each configuration. Finally, assuming rigid-wall boundary conditions, we find the Landau levels when the linear defect is itself magnetized. Remarkably, our analysis reveals that the Landau quantization occurs even in the absence of gauge fields provided the string is endowed with spin.
\end{abstract}
%
%
\maketitle
\section{Introduction}
In the last decade, a renewed interest in cosmic strings has been witnessed after a period of ostracism \cite{vilenkin,kibble,simanek,mota}. Cosmic strings are hypothetical massive objects that may have contributed, albeit marginally, to the anisotropy of the cosmic microwave background radiation and, consequently, to the large scale structure of the universe \cite{linde,ringeval}.  Actually, their existence is also supported in superstring theories with either compactified or extended extra dimensions. Both static and rotating cosmic strings can be equally responsible for some remarkable effects such
as particle self-force \cite{Eugenio1,CelioSelf} and gravitational lensing \cite{Sazhin}, as well as for production of highly energetic particles \cite{economou,valdir2,svaiter}.

Rotating cosmic strings, as well as their static counterparts, are one-dimensional stable topological defects probably formed during initial stages of the universe. They are characterized by a wedge parameter $\alpha$ that depends on its linear mass density, $\mu$, and by the linear density of angular momentum $J$. Initially, they were described as general relativistic solutions of a Kerr spacetime in (1+2) dimensions \cite{deser}, and then naturally extended to the four-dimensional spacetime  \cite{gott}. Notably, out of the singularity, cosmic strings (static or rotational) present a flat spacetime geometry with some remarkable global properties. These properties include theoretically predicted effects such as gravitomagnetism and (non-quantum) gravitational Aharanov-Bohm effect \cite{valdir3,Valdir4}.

Cosmic string may eventually present an internal structure \cite{jensen} generating a Goedel spacetime featuring an exotic region which allows closed time-like curves (CTC's) around the singularity. The frontier of this region is at a distance proportional to $J/\alpha$ from the string, thus offering
a natural boundary condition. Rotating cosmic strings were also studied in the Einstein-Cartan theory \cite{soleng,ozdemir} and in teleparallel gravity \cite{andrade}, in which the region of CTC's  was examined. There are also studies of these objects in the extra-dimensional context including their
causal structure, which raised criticisms on the real existence of the CTC's region \cite{slagter}.

Regarding Landau levels, in the spacetime of a stationary spinning cosmic string one does not find much literature \cite{mosta,Marcony} in contrast to what happens with static strings (see \cite{valdir,knutt1,knutt2,eugenio}, and references therein). This is probably due to the analogies and possible technological applications \cite{bueno} found in condensed matter physics (e.g. disclination in crystals). It is precisely this gap what motivates our paper. Thus, to make some progress in this direction, we will present a fully relativistic study of a massive charged particle
coupled to a gauge field in the spacetime spanned by a rotating string, with the eventual addition of scalar potentials.

On top of worthing the mathematical challenge on its own, it is phenomenologically meaningful to assess such a calculation for a static magnetic field parallel to the cosmic string and then compare the outcome with the static string results found in the literature \cite{valdir}. It is also opportune to check the non-relativistic limit in order to improve a previous non-relativistic calculation made with a much simpler approach \cite{Marcony}.

After such an outset, we will examine the problem when cylindric scalar potentials of coulombian and linear types are also considered. Phenomenologically, the coulombian potential is associated with a self-force acting on a charged particle in the spacetime of a cosmic string \cite{Linet,Soura}, and the linear term represents a cylindric harmonic oscillator of confining nature. Finally, we will consider the rotating string endowed with an internal magnetic flux and will discuss the raising of the Landau quantization from a pure spacetime rotation.

From the astrophysical point of view, the motivation to the present analysis lies on the possibility of existing scenarios in which charged relativistic particles interact with cosmic strings in the presence of intergalactic magnetic fields, with transitions between the energy levels yielding a spectrum that allows not only to identify a cosmic string, but to differentiate a static string from a rotating one. Such scenarios would also allow getting a reasonable estimate of the angular momentum of the string and, as a consequence, of the size of its CTCs frontier. Indeed, we will do so at the end of the paper.

The paper is organized as follows: In section II, we obtain the exact energy eigenvalues of the Klein-Gordon equation in the metric of a stationary rotating cosmic string coupled to a static magnetic field. In section III, we solve the problem along with some additional external potentials. In section IV, we consider a rotating string with an internal magnetic flux. Finally, in section V we conclude with some remarks.


\section{Spinless charged particle in a rotating cosmic string spacetime surrounded by an external magnetic field}

To start, we shall consider a massive, charged, relativistic spinless quantum particle in the spacetime of an idealized stationary rotating cosmic string. It means that the string has no structure and its metric is given by \cite{mazur}
\begin{equation}\label{01}
ds^2=c^2dt^2+2acdtd\phi-(\alpha^2\rho^2-a^2)d\phi^2-d\rho^2-dz^2,
\end{equation}
where the string is placed along the $z$ axis and the cylindrical coordinates are labeled by $(t, \rho, \phi,z)$ with the usual ranges.
Here, the rotation parameter $a=4GJ/c^3$ has units of distance and $\alpha=1-4\mu G/c^2$ is the wedge parameter which determines the angular
deficit, $\Delta \phi=2\pi(1-\alpha)$, produced by the cosmic string. The letters $c, G$ and $\mu$ stand for the light's speed, gravitational Newton's constant, and mass linear density of the string.

In order to investigate the relativistic quantum motion in the presence of a gauge potential and in a curved spacetime, let us consider the Klein-Gordon equation whose covariant form is written as 
\begin{equation}\label{02}
\left[\frac{1}{\sqrt{-g}}
D_{\mu} \left(\sqrt{-g}g^{\mu\nu}D_{\nu} \right) + \frac{m^2c^2}{\hbar^2}\right]\Psi=0,
\end{equation}
%
where $D_{\mu}=\partial_\mu - \frac{ie}{\hbar c}A_{\mu}$,  $e$ is the electric charge
and $m$ is the mass of the particle; $\hbar$ is as usual the Planck constant,
$g^{\mu\nu}$ is the metric tensor  and $g = \det g^{\mu\nu}$.
Assuming the existence of a homogeneous magnetic field $B$ parallel to the string,
the vector potential can be taken as $\vec{A}=(0, A_{\phi},0)$, with $A_\phi=\meio \alpha B\rho^2$.

The cylindrical symmetry of the background space, given by Eq. (\ref{01}), suggests
the factorization of the solution of  Eq. (\ref{02}) as
\beq \label{ansatz}
\Psi(\rho,\phi,z;t)=e^{-i\frac{E}{\hbar}t} e^{i(\ell\phi+k_zz)} R(\rho),
\eeq
where $R(\rho)$ is the solution of the radial equation given by
\begin{eqnarray} \label{eqR}
\frac{d^2R}{d\rho^2}+\frac{1}{\rho}\frac{dR}{d\rho}-\Lambda\,\frac{R}{\rho^2}-
\frac{e^2B^2}{4\hbar^2c^2} {\rho^2} R + \Delta \, R=0,
\end{eqnarray}
with
\bea
\Lambda &=& \left(\frac{\ell}{\alpha}+\frac{a E}{\alpha\hbar c}\right)^2 \label{eqL},\\
\Delta &=& \frac{E^2}{\hbar^2 c^2}-\frac{m^2c^2}{\hbar^2}-k_z^2+\frac{e B}{\hbar c}
\left(\frac{\ell}{\alpha}+\frac{a E}{\hbar c\alpha}\right);\label{eqD}
\eea
$k_z$ and $E$ are $z$-momentum and energy of the particle, and
 $\ell$  the azimuthal angular quantum number.
The solutions of Eq. (\ref{eqR}) can be found by means of the following transformation
\beq
R(\rho) = \exp{\left(-\frac{B e\rho^2}{4\hbar c}\right)}~\rho^{\sqrt{\Lambda}} F(\rho).
\eeq
 Substituting the above expression in Eq.(\ref{eqR}) we obtain
\bea \label{04}
\rho F''(\rho) + \left(1+2\sqrt{\Lambda}-\frac{B e}{\hbar\,c}\, \rho^2\right) F'(\rho)+
\left[\Delta-\frac{B e}{\hbar\,c}\left(1+\sqrt{\Lambda}\right)\right] \rho \,F(\rho) = 0.
\eea
Now, let us consider the change of variables $z=({Be}/{2\hbar c})\rho^2$. Thus, Eq. (\ref{04})
assumes the familiar form
\beq \label{eqHeun}
z F''(z) + \left(\sqrt{\Lambda}+1-z\right) F'(z)-\left[\frac{1}{2}\left(\sqrt{\Lambda}+1\right)-
\frac{\hbar c}{2 e B}\Delta\right]F(z)=0,
\eeq
which is the wellknown confluent hypergeometric equation,
whose linearly independent solutions are
\bea
F^{(1)}(z) &\!=\!& {}_1F_1\!\left(\frac{1}{2}+\frac{\sqrt{\Lambda}}{2}-
\frac{\hbar c}{2 e B}\Delta;\sqrt{\Lambda}+1;z\right),\\
F^{(2)}(z) &\!=\!& z^{-\sqrt{\Lambda}}\,{}_1F_1\!\left(\frac{1}{2}-\frac{\sqrt{\Lambda}}{2}-
\frac{\hbar c}{2 e B}\Delta;1-\sqrt{\Lambda};z\right).
\eea
Therefore, the radial solutions, $R(\rho)$, can be written as
\bea
R^{(1)}(\rho) &\!=\!& A_1\exp\left(\!-\frac{Be\rho^2}{4\hbar c}\right)\rho^{\sqrt{\Lambda}}\,{}_1F_1\!\left(\frac{1}{2}+\frac{\sqrt{\Lambda}}{2}-
\frac{\hbar c}{2 e B}\Delta;1+\sqrt{\Lambda};\frac{Be\rho^2}{2\hbar c}\right)\\
R^{(2)}(\rho) &\!=\!& A_2\exp\left(\!-\frac{Be\rho^2}{4\hbar c}\right)\rho^{-\sqrt{\Lambda}}\,{}_1F_1\!\left(\frac{1}{2}-\frac{\sqrt{\Lambda}}{2}-
\frac{\hbar c}{2 e B}\Delta;1-\sqrt{\Lambda};\frac{Be\rho^2}{2\hbar c }\right)
\label{radialfunction}
\eea
where $A_1$ e $A_2$ are normalization constants. The second solution is not physically
acceptable at the origin and we discard it.
Because confluent hypergeometric function diverge exponentially when $\rho\rightarrow\infty$,
in order to have asymptotically acceptable physical solutions we have to impose the condition
\beq
\frac{1+\sqrt{\Lambda}}{2}-\frac{\hbar c}{2 e B}\Delta=-n, \label{eqn}
\eeq
where $n$ is a positive integer. Substituting $\Lambda$ and $\Delta$
given by Eqs. (\ref{eqL}) and (\ref{eqD}), respectively, into Eq. (\ref{eqn}), we obtain
the following result
\bea \label{enq}
\frac{E^2} {B e \hbar c} + \left(\frac{a E}{\hbar c \alpha}+ \frac{\ell}{\alpha} \right ) -
\left \vert \frac{a E}{\hbar c \alpha}+\frac{\ell}{\alpha} \right \vert - \frac{c}{Be\hbar}
(\hbar^2 k^2+m^2 c^2)-\frac{1}{4}\frac{Bea^2}{\hbar c^3}=2n+1,
\eea
from which we can read the energy eigenvalues as
\bea
E_{n,\ell}=\frac{Bea}{2\alpha}~\frac{\left\vert\ell\right\vert -\ell}{\ell} \pm \sqrt{m^2c^4 \!+\! k^2\hbar^2c^2 \!+\! \left(\frac{Bea}{2\alpha}~\frac{\left\vert\ell\right\vert -\ell}{\ell}\right)^{\!\!2} \!\!+
\! {B\hbar c e} \left(\!2n+1 +\frac{|\ell|}{\alpha}-\frac{\ell}{\alpha}\right)}.\label{eqEnl}
\eea
This expression shows that the energy eigenvalues are not invariant under the interchange of positive and negative eigenvalues of the azimuthal quantum number $\ell$. This is a consequence of the spacetime topological twist around the spinning string which now depends not only on $\alpha$ but also on $a$ (see Eq.(\ref{01}) ). It is worth noticing that by turning off the string rotation, i.e. making $a=0$, we obtain an already known expression \cite{eugenio} valid for the static string. Notice also that for positive $\ell$, the energy spectra of both static and rotating strings are identical.

\subsection*{Non-relativistic limit}

The non-relativistic expression can be attained by considering $E^2/c^2-m^2c^2\approx 2mE$
in the previous equation. In this case, Eq. (\ref{eqEnl}) turns into
\bea
E_{n,\ell}\approx\frac{1}{1+\frac{eBa}{2mc^2 \alpha}(1-|\ell|/\ell)}
\left[\frac{\hbar^2 k^2}{2m}+\frac{Be\hbar}{2mc} \left(2n+1+
\frac{|\ell|}{\alpha}-\frac{\ell}{\alpha}\right) \right].
\eea
As a result, we can see that for $\ell >0$ (i.e. particle orbiting parallel to the string rotation)
the energy levels are the same for both static \cite{valdir} and spinning strings.
Otherwise, for antiparallel orbits ($\ell < 0$),
the allowed spectrum depends on the angular momentum density of the string (recall that $a=4GJ/c^3$).

In this case, if we consider the slow rotation approximation, where the terms $\mathcal{O}(a^2)$
are neglected, we have
\beq
\Delta E_{n,\ell}/E^{(0)}_{n,\ell}\approx -eBa/\alpha m c^2
\eeq
where $\Delta E_{n,\ell}$ is the relative difference of our result compared
to $E^{(0)}_{n,\ell}$, for the static string
levels \cite{valdir}. This result improves the one found in \cite{Marcony}
where further approximations were made.

\section{Cylindrically symmetric scalar potential in a rotating cosmic string spacetime surrounded by an external magnetic field}

In this section we shall perform a generalization of the analysis above done,
through the addition of the following cylindrically symmetric scalar potential  \cite{SCHAFER,eugenio},
\beq
S(\rho)= \frac{\kappa}{\rho}+{\nu}\,{\rho},	\label{scalar}
\eeq
where $\kappa$ and $\nu$ are constants.
	
In order to consider the influence of this potential on the quantum dynamics of the particle,
we have to modify Eq. (\ref{02}) by adding Eq. (\ref{scalar})  to the mass term
in such a way that $\frac{mc}{\hbar}$ is replaced by $\frac{mc}{\hbar} + S(\rho)$. Thus,
introducing this modification into Eq. (\ref{02}) and considering the ansatz
given by Eq.(\ref{ansatz}), we obtain the following radial equation
\begin{eqnarray} \label{eqRmod}
\frac{d^2R}{d\rho^2}+\frac{1}{\rho}\frac{dR}{d\rho}-\mathfrak{L}\,\frac{R}{\rho^2} -
2{M}\kappa\,\frac{R}{{\rho}}- 2{M}\nu\rho\,{R} - \Omega^2 {\rho^2} R + \mathfrak{D} \, R=0,
\end{eqnarray}
where
\bea \label{eqB}
{M}&=&\frac{mc}{\hbar}\\
\Omega^2 &=&M^2\omega^2+\nu^2\\
\mathfrak{L} &=& \left(\frac{\ell}{\alpha}+\frac{a}{\alpha}\mathcal{E}\right)^2
+\kappa^2\label{eqLb}\\
\mathfrak{D} &=& \mathcal{E}^2+2M\omega\left(\frac{\ell}{\alpha}+\frac{a}{\alpha}\mathcal{E}\right)
-{M}^2-2\kappa\nu-k_z^2, \label{eqDd}
\eea
 $2M\omega={e B}/{\hbar c \alpha}$ and $\mathcal{E}=E/\hbar c$. For convenience,
 let us define a new funtion $H(\rho)$ such that
\bea \label{transf}
R(\rho) = \exp\!\left({-\frac{1}{2} \Omega \rho^2 - \frac{M\nu}{\Omega}\rho}\right) \,
\rho^{\sqrt{\mathfrak{L}}}\, H(\rho).
\eea
Thus, using the redefinition $\sqrt{\Omega}\rho \rightarrow \rho$, Eq. (\ref{eqRmod}) reads
\bea \label{heunbsol}
\frac{d^2 H}{d\rho^2}&+&\left(\frac{1+2\sqrt{\mathfrak{L}}}{\rho}-
\frac{2M\nu}{\Omega^{3/2}}-2\rho\right) \frac{d H}{d\rho}\nonumber \\&+&
\left[ \frac{M^2\nu^2}{\Omega^3}+\frac{\mathfrak{D}}{\Omega} -
2\sqrt{\mathfrak{L}}-2-\frac{1}{2}\left(\frac{4M\kappa}{\sqrt{\Omega}}
+(1+2\sqrt{\mathfrak{L}})\frac{2M \nu}{\Omega^{3/2}}\right)\frac{1}{\rho} \right]H=0.
\eea
which corresponds to the biconfluent Heun equation \cite{ARRIOLA, RONVEAUX}.
Written in the standard form
\bea \label{heunB}
{H_{b}}''(z)+\left(\frac{1+\alpha}{z}-\beta-2z\right){H_{b}}'(z)+
\left[\gamma-\alpha-2-\frac{1}{2}[\delta+(1+\alpha)\beta]\frac{1}{z} \right] {H_{b}}(z)=0,
\eea
its solutions are the so-called biconfluent Heun functions
\beq
{H_{b}}(z)=\textit{C}_1 {H_{b}}(\alpha,\beta,\gamma, \delta;z)+
\textit{C}_2 \, z^{-\alpha} {H_{b}}(-\alpha,\beta,\gamma, \delta;z),
\eeq
with $C_1$ and $C_2$ being normalization constants. If $\alpha$ is not a
negative integer, the biconfluent Heun functions can be written as \cite{DECARREAU,VIEIRA}
\beq
{H_{b}}(\alpha,\beta,\gamma, \delta; z)= \sum^{\infty}_{j=0} \frac{A_j}{(1+\alpha)_j}\frac{z^j}{j!}
\eeq
where the coefficients $A_j$ obey the three-terms recurrence relation ($j\geq 0$)
\beq
A_{j+2}=\left[(j+1)\beta+\frac{1}{2}[\delta+(1+\alpha)\beta]\right] A_{j+1}
-(j+1)(j+1+\alpha) (\gamma-\alpha-2-2j) A_j
\eeq
Comparing directly Eqs. (\ref{heunbsol}) and (\ref{heunB}), we obtain the
following analytical solutions for $H(\rho)$
\bea \label{eqHb1}
H^{(1)}(\rho)&=&\textit{c}_1 {H_{b}}\left(2\sqrt{\mathfrak{L}},\frac{2M \nu}{\Omega^{3/2}},\frac{M^2\nu^2}{\Omega^{3}}
+\frac{\mathfrak{D}}{\Omega}, \frac{4M\kappa}{\sqrt{\Omega}};\sqrt{\Omega}\rho\right)\\
H^{(2)}(z)&=&\textit{c}_2\, \rho^{-2\sqrt{\mathfrak{L}}} {H_{b}}
\left(-2\sqrt{\mathfrak{L}},\frac{2M \nu}{\Omega^{3/2}},\frac{M^2\nu^2}{\Omega^{3}}+
\frac{\mathfrak{D}}{\Omega}, \frac{4M\kappa}{\sqrt{\Omega}};\sqrt{\Omega}\rho\right) \label{eqHb2}
\eea
where we have substituted back $\rho\rightarrow \sqrt{\Omega}\rho$ in the above expressions.
In view of Eq. (\ref{transf}) and the fact that the solution given by Eq. (\ref{eqHb2}) is
divergent at the origin, we will cast it off. Moreover, the biconfluent Heun functions are
highly divergent at infinity and so we need to focus on their polynomial forms.
Indeed, the biconfluent Heun function becomes a polynomial of degree $n$ if the
 following conditions are both satisfied (see \cite{VIEIRA} and references therein),
\bea
\gamma-\alpha-2&=&2n,~~  n=0,1,2,...\label{En}\\
A_{n+1}&=&0,
\eea
where $A_{n+1}$ has $n+1$ real roots when $1+\alpha>0$ and $\beta \in \mathbb{R}$.
It is represented as a three-diagonal $(n+1)$-dimensional determinant, namely,
\begin{eqnarray}
\hskip .truecm \left|\!\!
\begin{array}{ccccccc}
\delta' \! & 1  & 0 & 0 \! & \!\ldots & \!\ldots & 0\\
2 (1\!+\!\alpha)n   & \delta' \!-\!\beta  & 1 & 0 \! & \!\ldots & \!\ldots & 0  \\
0 & 4(2\!+\!\alpha)(n\!-\!1) & \delta' \!-\!2\beta & \!1\!& 0 & \ldots & 0 \\
0 & 0 & \gamma_2 &\! \delta'\!-\!3\beta \! &1 & \ldots &\vdots \\
\vdots & \vdots & 0 & \!\ddots\! & \ddots & \ddots & 0 \\
\vdots & \vdots & \vdots & \!\vdots\! & \gamma_{j-1} & ~~\delta'_{s-1} & 1\\
0 & 0 & 0 & 0 & 0 & 	\gamma_s & \delta'_s
\end{array}
\!\!\right|\!=0,\hskip .truecm
\label{DeltaN}
\end{eqnarray}
where
\bea
\delta'=-\frac{1}{2}[\delta+(1+\alpha)\beta]\\
\delta'_s=\delta'-(s+1)\beta\\
\gamma_s=2(s+1)(s+1+\alpha)(n-s). \label{eqC}
\eea
As an important consequence of Eq. (\ref{En}), we have
\beq \label{eq2n}
\frac{M^2\nu^2}{\Omega^3}+\frac{\mathfrak{D}}{\Omega}-2\sqrt{\mathfrak{L}}-2 = 2n,
\eeq
which means that the energy eigenvalues obey a quantization condition.
Differently from Eqs. (\ref{eqn}) and (\ref{enq}), now
 we have a fourth order expression for the energy which is given by
\beq \label{hugeE}
D_4 \mathcal{E}^4+D_3 \mathcal{E}^3+D_2 \mathcal{E}^2+D_1 \mathcal{E} + D_0 =0,
\eeq
where
\bea
D_4 &=& \frac{1}{\Omega^2}\nonumber\\
D_3 &=& \frac{4M\omega}{\Omega^2}\frac{a}{\alpha}\nonumber\\
D_2 &=& \frac{2 M^2\nu^2}{\Omega^4}-\frac{4(n+1)}{\Omega}+
\frac{2}{\Omega^2}\left(L+ {2M^2\omega^2}\frac{a^2}{\alpha^2}\right)\label{c2} \\
D_1 &=& \left[\frac{2 M^2\nu^2}{\Omega^4}-\frac{4(n+1)}{\Omega}+
\frac{2L}{\Omega^2}\right]{2M\omega}\frac{a}{\alpha}
-\frac{8a}{\hbar c \alpha}\frac{\ell}{\alpha} \nonumber\\
D_0 &=& \frac{M^2\nu^2}{\Omega^3}\left[\frac{M^2\nu^2}{\Omega^3}-4(n+1)\right]
+\left[\frac{2 M^2\nu^2}{\Omega^3}-4(n+1)+\frac{L}{\Omega}\right]\frac{L}{\Omega}\nonumber\\
&+& 4(n+1)^2-\frac{4\ell^2}{\alpha^2}-4\kappa^2,\nonumber
\eea
with $L=2M\omega\frac{\ell}{\alpha}-M^2-2\kappa\nu -k^2_z$.
Unfortunately, the analytical solutions for the energy eigenvalues are given by huge (algebraic) expressions.
However, we can manage them in some particular cases which will be presented in what follows.

\subsection{The rotation vanishes (${\bf a=0}$)}
In this case, we obtain the following result for the energy eigenvalues
\beq
E/\hbar c=\pm\left[k^2_z+ \frac{M^4 \omega^2}{\nu^2+M^2 \omega^2}+2\kappa\nu
-2M\omega \frac{\ell}{\alpha}+2\,\Omega\!\left(\!n+1+
\sqrt{\frac{\ell^2}{\alpha^2}+\kappa^2} \right)\! \right]^{\!\frac{1}{2}}\!\!,
\eeq
which coincides with the one already obtained in the literature \cite{eugenio}.

\subsection{The rotation vanishes and there is no scalar potential (${\bf a=0}$, ${\bf \kappa=0}$, ${\bf \nu=0}$)}
In the present situation, we have that $\Omega=M\omega$ and then the energy eigenvalues are given by
\beq
E/\hbar c=\pm\left[k^2_z+ {M^2}+2M\omega\!\left(\!n+1+{\frac{|\ell|}{\alpha}}
-\frac{\ell}{\alpha} \right)\! \right]^{\!\frac{1}{2}}\!\!.
\eeq
However, in this case the biconfluent Heun solution does not have the odd terms
as we can see expanding Eq. (\ref{eqHb1}) or from Eqs. (\ref{DeltaN})-(\ref{eqC}).
Therefore, the above expression only make sense when we consider the even terms,
or equivalently when $n\rightarrow 2n$ \cite{valdir}. Another way to see this is verifying that
\bea
{H_{b}}(2\sqrt{\Lambda}, 0, \frac{\Delta}{M\omega}, 0, \sqrt{M\omega}\rho) =
{}_2F_1\!\left(\frac{1+\sqrt{\Lambda}}{2}-\frac{\Delta}{4M\omega}, 1+\sqrt{\Lambda},
\sqrt{M\omega}\rho^2 \right),
\eea
and thus, showing the correspondence between conditions (\ref{eqn}) and (\ref{eq2n})
in this particular case.

\subsection{Linear confinement ($\kappa=0$)}
In this case, the Coulomb-type potential term is absent, and as a consequence
the scalar potential is reduced to the linear term in
$\rho$. Thus, the solutions are now given by
\bea \label{conf1}
H^{(1)}(\rho)&=&\textit{c}_1 {H_{b}}\left(2\sqrt{{\Lambda}},\frac{2M \nu}{\Omega^{3/2}},\frac{M^2\nu^2}{\Omega^{3}}
+\frac{\Delta}{\Omega}, 0;\sqrt{\Omega}\rho\right)\\
H^{(2)}(z)&=&\textit{c}_2\, \rho^{-2\sqrt{{\Lambda}}} {H_{b}}\left(-2\sqrt{{\Lambda}},\frac{2M \nu}{\Omega^{3/2}},\frac{M^2\nu^2}{\Omega^{3}}+\frac{\Delta}{\Omega}, 0;\sqrt{\Omega}\rho\right) \label{conf2}
\eea
Again we discard the second solution because it diverges at $\rho=0$.
The condition to get polynomial solutions is now
\beq \label{eq2nd}
\frac{M^2\nu^2}{\Omega^3}+\frac{\Delta}{\Omega}-2\sqrt{{\Lambda}}-2 = 2n
\eeq
As before, the above condition implies in the quantization of the energy
eigenvalues which is equivalent to Eq. (\ref{hugeE}), with the coefficients
given by (\ref{c2}), with $\kappa=0$.
\section{Spinless particle in the rotating cosmic string spacetime with an internal magnetic flux}

We will now examine the relativistic Landau levels of a charged spinless particle in the spacetime
of a magnetized rotating string (namely, endowed with some intrinsic magnetic flux $\Phi$) with no external electromagnetic field \cite{furtadomoraes,passosfurtadomoraes}. The corresponding gauge coupling is obtained by making $B\rightarrow B = \Phi/\alpha \pi \rho^2$ in Eq. (4).
In this case, the radial equation reads
\begin{eqnarray}\label{eqRflux}
\rho^2\frac{d^2R}{d\rho^2}+{\rho}\frac{dR}{d\rho}+(\delta\rho^2-\Sigma) R=0. \end{eqnarray}
where $\Sigma$ and $\delta$ are given by
\bea \label{omega}
\Sigma &=& \left(\frac{\ell}{\alpha}+\frac{a}{\alpha}\mathcal{E}-
\frac{\epsilon \Phi}{\alpha}\right)^{\!\!2}\\
\delta &=& {\mathcal{E}^2}-M^2-k_z^2
\eea
with $\epsilon = e/2\pi\hbar c$.\\
The solutions of Eq. (\ref{eqRflux}) are written in terms of Bessel's functions
of the first kind, $J_{\lambda}(z)$, and second kind, $Y_{\lambda}(z)$, as
\bea
R(\rho) = C_1\,J_{\sqrt{\Sigma}}\left(\sqrt{\delta}\,\rho\right)+
C_2 Y_{\sqrt{\Sigma}} \left(\sqrt{\delta}\rho\right),
\eea
with $C_1$ and $C_2$ being constants. The function $J_\lambda (z)$ is
different from zero at the origin when $\lambda=0$.
Otherwise, $Y_{\sqrt{\Sigma}}$ is always divergent at the origin.
Thus, we will discard it and consider $\lambda\neq0$.
It is worth pointing out that when $\Phi=0$, we reobtain the wave function found in \cite{Krori}. To find the energy eigenvalues, we will impose the so called hard-wall condition. With this boundary condition, the wave function of the particle vanishes at some $\rho=r_w$ which is an arbitrary radius far away from the origin. Thus, we can use the asymptotic expansion for large arguments of $J_\lambda (z)$, given by
\begin{equation}
J_{\lambda}(z)\approx \sqrt{\frac{2}{\pi z}}
\cos{\left(z-\frac{\lambda\pi}{2}-\frac{\pi}{4}\right)}, \label{bessel}
\end{equation}
from which we obtain
\begin{equation}\label{condition}
\sqrt{\delta}r_w-\frac{\sqrt{\Sigma}\pi}{2}-\frac{\pi}{4}=\frac{\pi}{2}+ n\pi,
\end{equation}
for $n\in\mathbb{Z}$. Substituting Eqs. (\ref{omega}) and (50) into (\ref{condition}), we get
\begin{equation}
r_\omega\sqrt{\mathcal{E}^2-M^2-k^2_z}\mp\frac{\pi}{2}
\left( \frac{\ell}{\alpha}+\frac{a}{\alpha}\,\mathcal{E}-
\frac{\epsilon\Phi}{\alpha}\right)=\left(n+\frac{3}{4}\right)\pi,
\label{potencias}
\end{equation}
where the upper and lower signals correspond to $\ell/\alpha + a\mathcal{E}/\alpha-\epsilon\Phi/\alpha\leq 0$
or $\ell/\alpha + a\mathcal{E}/\alpha-\epsilon\Phi/\alpha>0$, respectively. Equation (\ref{potencias}),
can be rewritten as the following second order equation
$$
A_1\,\mathcal{E}^2+A_2\,\mathcal{E}+A_3=0,
$$
with
\bea
A_1&=&r^2_\omega-\frac{a^2\pi^2}{4\alpha^2} \nonumber\\
A_2&=& -\frac{a\pi^2}{2\alpha}\left[\left(\frac{\ell}{\alpha}-
\frac{\epsilon \Phi}{\alpha}\right)\pm\left(2n+\frac{3}{2}\right)\right]\\
A_3&=-&r^2_\omega \left(M^2+k^2_z\right)-\left(\frac{\ell}{\alpha}-
\frac{\epsilon \Phi}{\alpha}\right)\frac{\pi^2}{4}-\left(n+
\frac{3}{4}\right)^2 \pi^2 \mp \left(n+\frac{3}{4}\right)\left(\frac{\ell}{\alpha}-
\frac{\epsilon\Phi}{\alpha}\right)\pi^2 \nonumber
\eea
Since $r_w $ is very large $\mathcal{E}$ reduces to
\bea
\mathcal{E}_{+} &\approx& + \sqrt{M^2+k^2_z} +\frac{a\pi^2}
{4\alpha r^2_\omega}\left[\frac{\ell}{\alpha}-
\frac{\epsilon \Phi}{\alpha} \pm \left(2n+\frac{3}{2}\right)\right] \label{E+}\\
\mathcal{E}_{-} &\approx& - \sqrt{M^2+k^2_z} +\frac{a\pi^2}
{4\alpha r^2_\omega}\left[\frac{\ell}{\alpha}-
\frac{\epsilon \Phi}{\alpha} \pm \left(2n+\frac{3}{2}\right)\right].\nonumber
\eea
Let us now hang up with $\mathcal{E}_{+} (=E_+/\hbar c)$ and assume that $k_z << M$.
Then, provided that $\ell/\alpha \geq \epsilon\,\Phi/\alpha$
(see Eq. (\ref{potencias}) ), we have
\beq
E_{+} \approx mc^2+\frac{a \pi^2 \hbar c}{4\alpha r^2_\omega}
\left[\frac{\ell}{\alpha}-\frac{\epsilon\,\Phi}{\alpha}+2n+\frac{3}{2}\right],
\eeq
which shows that in the absence of rotation, the energy eigenvalues reduce to
the rest energy of the particle irrespective of $\alpha$. In other words,
the eigenenergies are the same with or without the presence of a (static)
magnetized cosmic string in space  but split if the string rotates.

\section{Conclusions and Remarks}
We have analyzed the Landau levels of a spinless massive particle in the spacetime
of a rotating cosmic string by means of a fully relativistic approach.
Specifically, in Section II the Landau quantization has been derived in a static
and homogeneous magnetic field parallel
to the string by solving the covariant Klein-Gordon equation in the spacetime
of a conical singularity endowed with spin. 
The physically significant role played by the string rotation, as introduced into the metric,
becomes apparent in the particle's energy spectrum. As shown in Eqs.(\ref{radialfunction}) and
(\ref{eqEnl}) eigenvalues and eigenfunctions depend nontrivially on
both the string spinning parameter $a$, the topological deficit $\alpha$,
and the particle's angular momentum $l$.
Turning off the string rotation, makes the Landau levels to collapse to those of a
static string \cite{eugenio,valdir}, as expected.
The non-relativistic limit of the energies was also found and equally well
compared with the static case; the present result improves and corrects a previous
one obtained by means of a simpler approach \cite{Marcony}.

In Section III we obtained the spectrum of the particle when a gauge potential
together with a scalar one are present in the space around the rotating string.
We shown that the eigensates are given by biconfluent Heun functions,
which in their polynomial representation allowed finding a quantization condition
on the energy levels. The general expression can be analytically obtained but looks
rather huge, so we decided to exhibit just some special relevant cases
which indeed confirm the results already obtained in \cite{valdir,eugenio}.

We have also tackled the problem of a rotating cosmic string endowed with an
internal magnetic flux with a hard-wall boundary far away from
the source (see Eqs. (\ref{bessel}) -  (\ref{E+}) in Section IV).
The resulting eigenfunctions converge to those found in the literature when the magnetic
flux vanishes \cite{Krori}, as expected. It is noteworthy
that the Landau levels of the spinning string remain the same even when such
internal magnetic flux fades away; namely, when there is no gauge field inside nor around.
This can be interpreted as an induction of the Landau quantization from the sole rotational
condition of the defect.
It is interesting to compare this result with that of a rotating spherical
source in Kerr spacetime obtained in \cite{Konno}.

Finally, as a phenomenological byproduct of our results, it is possible to provide a reasonable
estimate of the angular momentum  of the rotating cosmic string, $J$.
Consider a proton orbiting with angular velocity $\Omega$
around the string  very close to the CTC's frontier.
Now, for $a\approx c/\Omega$ and $\Omega=\omega_c = e B/ 2\alpha mc$ with $B\sim10^{-6}$ G
(which is the value of currently observable intergalactic magnetic fields \cite{russel}),
we conclude that the CTC's frontier is at about $10^{11}$ m from the string, which
corresponds to $J\sim10^{47}$ kg m/s.
This value is compatible with the one presented in \cite{mazur}
when the upper limit of the photon mass, $10^{-16}$ eV, is taken into account \cite{alfred,Geova}.

As a future perspective, we intend to study the problem by considering a spinorial particle.

\section*{Acknowledgements} M. S. Cunha, C. R. Muniz, and V. B. Bezerra would like to thank
to Conselho Nacional de Desenvolvimento Cient\'ifico e Tecnol\'ogico (CNPq) for the partial support.

\end{document}